\documentclass[12pt]{article}
\usepackage[pdftex]{graphicx}
\usepackage{caption}
\usepackage{wrapfig}
\usepackage{latexsym}
\usepackage[pdftex]{color}
\usepackage{pifont}
\usepackage{epstopdf}
\usepackage{tikz}
\usepackage{mathtools}
\usetikzlibrary{decorations.pathmorphing,patterns}

\newcommand{\bmp}[2][t]{\begin{minipage}[#1]{#2}}
\newcommand{\emp}{\end{minipage}}

\usepackage{amsmath}
\usepackage{amssymb}
\usepackage{textcomp}
\usepackage{tcolorbox}
\usepackage[title]{appendix}

\usepackage{sidecap}

\usepackage[square,numbers,sort&compress]{natbib} 

\newcommand{\bblubox}[1]{\begin{tcolorbox}[colframe=blue!75!white,title=#1]}
\newcommand{\eblubox}{\end{tcolorbox}}

\begin{document}

\title{Membrane Thickness Sensitivity\\ of Avian Prestin: Implications} 
\author{Kuni H Iwasa}
\date{\small{National Institute for Deafness and Other Communication Disorders, \\ National Institutes of Health, Bethesda, MD}}
\maketitle
\abstract{
Avian prestin is sensitive to membrane thickness as much as mammalian prestin, which undergoes conformational transitions in membrane area and thereby drives length changes of the cylindrical cell body of outer hair cells. The membrane thickness dependence of mammalian prestin stems from changes in hydrophobic profile in conformational states, accompanied by changes in their membrane area. Even though such area changes are not detected for avian prestin, it nonetheless bends hair bundles of avian short hair cells.  Here it is suggested that the motile function of avian prestin can be based on conformational transitions involving shearing deformation of the membrane protein, which also leads to membrane thickness sensitivity.
}

\section*{Introduction}
Membrane thickness dependence of a membrane protein arises from difference in hydrophobic profile of its conformational states. Specifically for mammalian prestin, which undergoes changes in the surface area \cite{ai1999} coupled with charge transfer $q$ across the membrane, membrane thickness dependence was expected since the volume of the protein is conserved during conformational changes. 

Indeed, that is the case with mammalian prestin: A reduction of membrane thickness shifts the transition voltage of the protein in the positive direction and an increase in membrane thickness has an opposite effect \cite{Fang2010}. This observation is consistent with the expectation that a decreased hydrophobic thickness of the protein is associated with an increase in the surface area on hyperpolarization \cite{Bavi2021,Futamata2022}.

Membrane thickness dependence can be closely associated with surface area changes during conformational transitions, which result in changes in the hydrophobic profile associated with thickness changes. It can be also associated with changes in the contour length of the hydrophobic interface between the protein and membrane lipid.

The mode of motion, with which avian prestin is associated, differs from that of mammalian prestin. Avian prestin is associated with bending of hair bundles of avian short hair cells \cite{Beurg2013p}, which is quite different from length changes of cell body, which mammalian prestin drives \cite{a1987}. 

How can these observations provide a physical picture?  The present report is an attempt to address this issue.?

\section*{Membrane thickness dependence }
Chicken prestin shows considerable membrane thickness dependence accompanied by somewhat smaller charge transfer compared with mammalian prestin \cite{Izumi2011}. \\

\begin{table}[h]

\begin{center}
\begin{tabular}{r | r c c} \hline
& $V_\mathrm{pk} $ & q & \normalsize{thickness sensitivity}\\ 
& (mV) & ($e$) & (mV/\%) \\ \hline
gerbil & $-88 \pm 11$ & $0.73\pm 0.07$ & $2.7 \pm 0.2$  \\
platypus & $-56\pm 11$ & $0.79\pm 0.10$ & $4.8\pm 0.1$ \\
\color{blue} \textbf{chicken} &  $54\pm 11$ &  $0.35 \pm 0.12$ &  $8.9\pm 0.7$ \\
\hline
\end{tabular} \\
\caption{Here, Peak potential $V_\mathrm{pk}$, motor charge $q$, and thickness sensitivity. Membrane thickness sensitivity is expressed as voltage shifts per \% change in the linear capacitance. $e$: the electronic charge. Taken from Ref.\ \cite{Izumi2011}.}
\end{center}
\label{tab:data}
\end{table}%

Avian prestin has motile charge $q$ smaller than that of mammalian prestin. However, its membrane thickness dependence is larger than that of mammalian counterpart. Another significant difference in the operating voltage $V_{pk}$ can be attributed to rather depolarized membrane potential of avian hair cells \cite{TanFett2013}, which is associated large hair bundle current due to cellular turning of the avian ear \cite{CrawFett1981,Iwasa2022}.

\subsection*{Interpretation}
Membrane thickness sensitivity results from difference in interaction energy of a membrane protein, which undergoes conformational transitions, with membrane lipid.  

Let us assume a model membrane protein has two states $S_0$ and $S_1$. Assume that $S_0$  and $C_1$ has, respectively, length of circumference $L_0$ and $L_1=L_0+\Delta L$, at which it interacts with membrane lipid. Let $\mathcal{E}(d-d_0)$ be the energy per unit length due to hydrophobic mismatch in state $S_0$ with membrane lipid with thickness $d$, where $d_0$ is the characteristic hydrophobic thickness of this state. Let $d_1$ be the characteristic thickness of $S_1$. 

The energy $E_0$ and $E_1$ due to hydrophobic mismatch respectively for $S_0$ and for $S_0$ can be expressed by 
\begin{subequations}
\begin{align}
E_0(d)&=L_0\mathcal{E}(d-d_0),\\
E_1(d)&=(L_0+\Delta L)\mathcal{E}(d-d_1).
\end{align}
\end{subequations}
It is likely that the function $\mathcal{E}(x)$ is dominated by even powers of $x$ because it should increase with the absolute value of the variable $x$. The simplest case could be $\mathcal{E}(x)=ax^2$, where $a$ is a constant, but it likely includes higher-order terms.
The difference in energy between the two states due to hydrophobic incompatibility is $E_\mathrm{diff}(d)=E_1(d)-E_0(d)$. 

Now the effect of membrane thickness change $d\rightarrow d+\Delta d$. This change shifts the energy difference between the two states by 
\begin{align}
\Delta E=E_\mathrm{diff}(d+\Delta d)-E_\mathrm{diff}(d).
\label{eq:deltaE1}
\end{align}
This energy difference changes the relative weight of the conformation states, i.e. the conformational state of lower energy is favored. That leads to a shift of voltage dependence of prestin.

\begin{figure}[h]
\begin{center}
\includegraphics[width=0.8\linewidth]{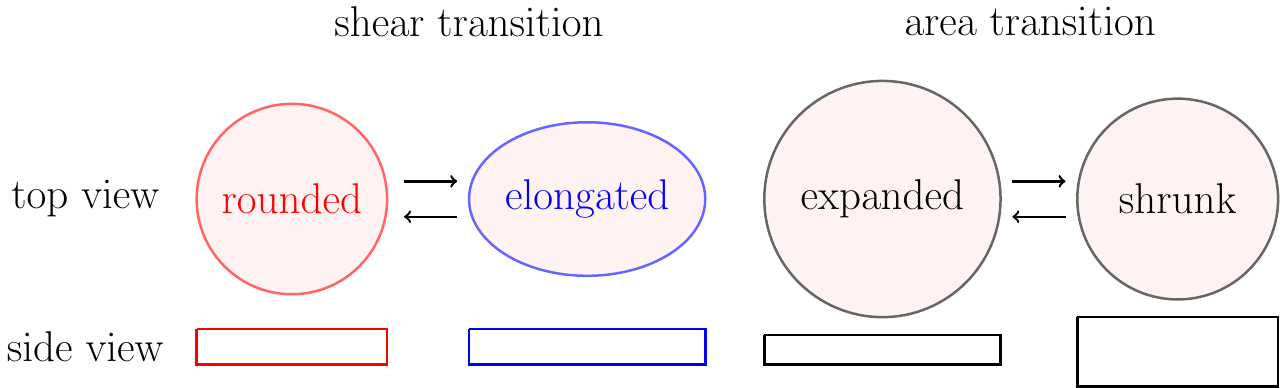}
\caption{Two types of conformational transitions, which result in membrane thickness dependence. Shear transitions (left) undergoe without membrane area changes. A transition from a circle to an ellipse, for example, changes the circumference length from $L_0$ to $L_0+\Delta L$ without changing the area. Area transitions (right) undergo change in the membrane area, which result in thickness change, affecting hydrophobic mismatch.}
\label{fig:deform}
\end{center}
\end{figure}

If we can assume that $\Delta d/d\ll 1$ but $\Delta L/L$ may not be small, Eq.\ \ref{eq:deltaE1} can be approximated by the first-order terms of expansion, i.e.
\begin{align}
\Delta E/\Delta d=L_0(\mathcal{E}'(d-d_1)-\mathcal{E}'(d-d_0))+\Delta L\mathcal{E}'(d-d_1)
\label{eq:deltaE2}
\end{align}
where $\mathcal{E}(x)'$ is the first derivative of $\mathcal{E}(x)$ with respect to $x$. In the simplest case $\mathcal{E}(x)=ax^2$, the first term is proportional to $d_1-d_0$. This special case illustrates that the first term in Eq.\ \ref{eq:deltaE2}  is due to the difference in hydrophobic thickness of the two states. The second term is due to the difference $\Delta L$ in the circumference in the two conformational states.

For mammalian prestin, which includes membrane area changes during conformational transitions \cite{ai1999}, changes in the hydrophobic thickness is expected by assuming volume conservation. In addition, changes in the circumference can be expected as schematically illustrated in Fig.\ \ref{fig:deform}.

Eq.\ \ref{eq:deltaE2} also shows that membrane thickness dependence, however, can arise from the difference $\Delta L$ in the circumference alone without changes in the surface area. If these membrane proteins are randomly oriented, membrane share should cancel each other and no collective motion is expected. This expectation is consistent with the observation that plasma membrane of HEK cells transfected with avian prestin did not show movement of cell membrane elicited by voltage pulses, unlike mammalian prestin \cite{TanHe2011}. That is because those avian prestin molecules expressed at low densities in the plasma membrane of those transfected cells are likely oriented randomly. The randomness of orientation of this protein does not lead to macroscopic movement.

\section*{Macroscopic shear}

\begin{figure}[h]
\begin{center}
\includegraphics[width=0.4\linewidth]{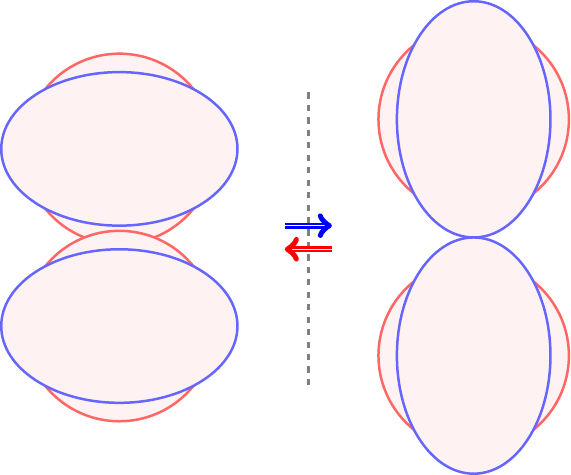}
\caption{A schematic representation of creating membrane shear. Molecular shear produced by conformational transitions can lead to membrane shear if the motile molecules are aligned. Two of the possible alignments are shown. The red and blue outlines illustrate two conformational states as  Fig.\ \ref{fig:deform} does. }
\label{fig:bend}
\end{center}
\end{figure}

Mean membrane shear can be achieved by shearing mode of a membrane protein as illustrated in Fig.\ \ref{fig:bend}. The main advantage of this mode compared with area motor mode is that bending shear can be provided without geometrical asymmetry of the cell membrane. This advantage, however, requires alignment of the motile molecules. 

\section*{Bending of hair bundle}

It has been found that avian prestin is also associated with motility \cite{Beurg2013p}. Instead of the cell body as mammalian prestin does, avian prestin drives hair bundles of short hair cells and thereby moves the tectorial membrane associated with those hair bundles \cite{Beurg2013p}.  

For the mechanism of this motion, localization of avian prestin is critically important. Histochemical study showed that prestin is located in the lateral plasma membrane \cite{Beurg2013p}. The authors point out in the appendix that avian short hair cells prominently expand toward the apex on the neural side and the cuticular place, which is associated with the hair bundle does not cover the neural side of the apical membrane \cite{Beurg2013p}. 

They propose that the neural side of the lateral membrane contracts on depolarization, tilting the hair bundle in the neural direction  \cite{Beurg2013p}. 

\section*{Discussion}

The mechanism proposed by Beurg et al.\ \cite{Beurg2013p} are based on the assumption that avian prestin undergoes membrane area changes, or ``area transitions,'' similar to mammalian prestin. This proposal is of interest even though the assumption contradicts experimental observations \cite{TanHe2011}. 

While the structural asymmetry can be confirmed by prestin-antibody staining, the asymmetry is not so large. It is unclear if relatively small asymmetry of the cell morphology as revealed by these images is sufficient to produce bending of the cell if the mode of the molecular displacement is area changes. 

Another issue is that the lateral membrane is in contact with the lateral membrane of another cell, unlike mammalian outer hair cells. A necessary  condition is that the neighboring cell is compliant in the direction of the movement. This requirement would apply to the submembranous cytoskeletal structure of those cells.

The neural side of the apical membrane appears to be the localization, which is logical for driving the hair bundle. However, antibody-staining does not show the presence of prestin. An analogous argument can be made with a molecular motor, which produces shear instead of membrane area changes.

Located in the lateral membrane, molecular ``shear generator'' could bend the hair bundle independent of the morphological asymmetry. However, generation of regional shear of the plasma membrane by molecular shear generators requires molecular alignment of the shear generators. Such alignment could be based on, e.g.\ cortical protein organization, and is plausible in view of the polarized organization of hair cells.

Those observations described here are of theoretical nature. To clarify the mechanism of hair bundle bending of avian short hair cells, further studies in both molecular and cellular levels are essential. Those studies concern details of localization and orientation of avian prestin on the cellular level. In addition, experimental determination of the upper bound of membrane area changes and studies of molecular structure of avian prestin would be critical.

\bibliography{/Users/kuni/Dropbox/wip/bib/ohc,/Users/kuni/Dropbox/wip/bib/tuning}

\begin{thebibliography}{11}
\providecommand{\url}[1]{\texttt{#1}}
\providecommand{\urlprefix}{ }

\bibitem[Adachi and Iwasa(1999)]{ai1999}
Adachi, M., and K.~H. Iwasa, 1999.
\newblock Electrically driven motor in the outer hair cell: Effect of a
  mechanical constraint.
\newblock \emph{Proc. Natl. Acad. Sci. USA} 96:7244--7249.

\bibitem[Fang et~al.(2010)Fang, Izumi, and Iwasa]{Fang2010}
Fang, J., C.~Izumi, and K.~H. Iwasa, 2010.
\newblock Sensitivity of prestin-based membrane motor to membrane thickness.
\newblock \emph{Biophys J} 98:2831--2838.

\bibitem[Bavi et~al.(2021)Bavi, abd Gustavo F~Contreras, Shen, Reddy, Milewski,
  and Perozo]{Bavi2021}
Bavi, N., M.~D.~C. abd Gustavo F~Contreras, R.~Shen, B.~G. Reddy, W.~Milewski,
  and E.~Perozo, 2021.
\newblock The conformational cycle of prestin underlies outer hair cell
  electromotility.
\newblock \emph{Nature} 600:553--558.

\bibitem[Futamata et~al.(2022)Futamata, Fukuda, Umeda, Yamashita, Tomita,
  Takahashi, Shikakura, Hayashi, Kusakizako, Nishizawa, Homma, and
  Nureki]{Futamata2022}
Futamata, H., M.~Fukuda, R.~Umeda, K.~Yamashita, A.~Tomita, S.~Takahashi,
  T.~Shikakura, S.~Hayashi, T.~Kusakizako, T.~Nishizawa, K.~Homma, and
  O.~Nureki, 2022.
\newblock Cryo-EM structures of thermostabilized prestin provide mechanistic
  insights underlying outer hair cell electromotility.
\newblock \emph{Nat Commun} 6208.

\bibitem[Beurg et~al.(2013)Beurg, Tan, and Fettiplace]{Beurg2013p}
Beurg, M., X.~Tan, and R.~Fettiplace, 2013.
\newblock A prestin motor in chicken auditory hair cells: active force
  generation in a nonmammalian species.
\newblock \emph{Neuron} 79:69--81.

\bibitem[Ashmore(1987)]{a1987}
Ashmore, J.~F., 1987.
\newblock A fast motile response in guinea-pig outer hair cells: the molecular
  basis of the cochlear amplifier.
\newblock \emph{J. Physiol.} 388:323--347.

\bibitem[Izumi et~al.(2011)Izumi, Bird, and Iwasa]{Izumi2011}
Izumi, C., J.~E. Bird, and K.~H. Iwasa, 2011.
\newblock Membrane thickness dependence of prestin orthologs: {T}he evolution
  of a piezoelectric protein.
\newblock \emph{Biophys. J} 100:2614--2622.

\bibitem[Tan et~al.(2013)Tan, Beurg, Hackney, Mahendrasingam, and
  Fettiplace]{TanFett2013}
Tan, X., M.~Beurg, C.~Hackney, S.~Mahendrasingam, and R.~Fettiplace, 2013.
\newblock Electrical tuning and transduction in short hair cells of the chicken
  auditory papilla.
\newblock \emph{J. Neurophysiol.} 109:2007 -- 2020.

\bibitem[Crawford and Fettiplace(1981)]{CrawFett1981}
Crawford, A.~C., and R.~Fettiplace, 1981.
\newblock An electrical tuning mechanism in turtle cochlear hair cells.
\newblock \emph{J. Physiol.} 312:377--412.

\bibitem[Iwasa(2022)]{Iwasa2022}
Iwasa, K.~H., 2022.
\newblock Of Mice and Chickens: Revisiting the RC Time Constant Problem.
\newblock \emph{Hearing Res.} 423:108422.

\bibitem[Tan et~al.(2011)Tan, Pecka, Tang, Okoruwa, Zhang, Beisel, and
  He]{TanHe2011}
Tan, X., J.~L. Pecka, J.~Tang, O.~E. Okoruwa, Q.~Zhang, K.~W. Beisel, and
  D.~Z.~Z. He, 2011.
\newblock From zebrafish to mammal: functional evolution of prestin, the motor
  protein of cochlear outer hair cells.
\newblock \emph{J Neurophysiol} 105:36--44.

\end{thebibliography}

\end{document}